\documentclass[twocolumn]{aastex631} 

\usepackage{wrapfig}
\usepackage{amsmath}
\usepackage{natbib}
\usepackage{graphicx}
\usepackage{hyperref}

\begin{document}

\title{SKYSURF VIII - Modeling SKYSURF Completeness Data for Comparison to the Hubble Space Telescope Exposure Time Calculator}

\author[0009-0004-8596-6433]{Zachary Goisman}
\affiliation{School of Earth and Space Exploration, Arizona State University, Tempe, AZ 85287-6004, USA}
\affiliation{School of Electrical and Computer Engineering, Arizona State University, Tempe, AZ 85287-6004, USA}

\author[0000-0001-6650-2853]{Timothy Carleton}\affiliation{School of Earth and Space Exploration, Arizona State University, Tempe, AZ 85287-6004, USA}

\author[0000-0003-3329-1337]{Seth H. Cohen}\affiliation{School of Earth and Space Exploration, Arizona State University, Tempe, AZ 85287-6004, USA}

\author[0000-0002-2099-639X]{Delondrae Carter}\affiliation{School of Earth and Space Exploration, Arizona State University, Tempe, AZ 85287-6004, USA}

\author[0000-0001-8156-6281]{Rogier A. Windhorst}\affiliation{School of Earth and Space Exploration, Arizona State University, Tempe, AZ 85287-6004, USA}

\author[0000-0003-3351-0878]{Rosalia O'Brien}\affiliation{School of Earth and Space Exploration, Arizona State University, Tempe, AZ 85287-6004, USA}

\author{Eyan Weissbluth}\affiliation{School of Earth and Space Exploration, Arizona State University, Tempe, AZ 85287-6004, USA}

\begin{abstract}
Accurately assessing image source completeness is critical for interpreting measurements on telescope data. Using the Wide Field Camera 3 (WFC3) and Advanced Camera for Surveys (ACS) data from the Hubble Space Telescope ($HST$) archival project ``SKYSURF", we model galaxy completeness as a function of the exposure time and background of an image. This is accomplished by adding simulated objects with varying magnitudes and sizes into these images, and determining the detection rate for each set of parameters. The fifty percent completeness results are then compared to the Exposure Time Calculator (ETC), in order to assess the differences between the STSCI ETC and our analysis of the archival data. Ultimately, we find that, for extended galaxies, the ETC predicts a 1-2 magnitudes fainter completeness limit than our data. We believe the difference is due to the ETC's flat surface brightness profiles being less accurate at predicting for extended sources compared to our more realistic profiles.
\end{abstract}


\section{Introduction}
SKYSURF aims to study the Extragalactic Background Light (EBL) and determine the sources of diffuse light \citep{Windhorst, Carleton,O'Brien,Driver}. Specifically, the EBL is all the UV-to-IR radiation from outside the Milky Way, encompassing the light from all other galaxies in the observable universe. Diffuse light is the component of the sky surface brightness, as seen by Hubble, whose origin is unknown; however, it potentially is a part of the EBL. Number counts, counting the number of sources above a flux threshold, are a key component in this study. In particular, faint and diffuse objects contribute to the EBL, though it is unclear how much they contribute to it. Regardless, determining the ability to detect sources of a given magnitude is critical to interpret the resulting number counts. 

We define completeness as the percentage of detected objects out of the total number of objects. The completeness simulations in \cite{Snigula} and \cite{Zaritsky} present a framework for measuring and analyzing the completeness of images with low object count through the use of inserted objects. We aim to extend this analysis to the entire Hubble archive using the SKYSURF project database in \cite{Windhorst} and \cite{Carter}. Since completeness measurements on the whole archive are not feasible, estimations by the $HST$ Exposure Time Calculator (ETC) serve as a viable alternative. However, the ETC was designed for idealized cases, and our process uses galaxies with realistic profiles, compared to the purely flat surface brightness profiles that the ETC uses. Therefore, our completeness measurements will determine how accurate these estimations truly are, and show where they need adjustments. Furthermore, we can use our obtained completeness models in the place of the ETC for future research. All magnitudes use the AB system from \cite{Oke}.

Our paper describes the used SKYSURF data in \S\ref{sec:data}, with methods of determining the completeness limits of images in \S\ref{sec:analysis}. These methods include inserting simulated objects of various parameters using GalSim in \S\ref{sec:inserted}, and the used Source Extractor parameters in \S\ref{sec:sourceExtractorAnalysis}. Using this background, we present the results in \S\ref{sec:results}. All the fitting results and ETC comparison plots are described in this section. Lastly, the results are discussed in \S\ref{sec:discussion}, a summary is in \S\ref{sec:summary}, and acknowledgments are in \S\ref{sec:acknowledgements}.

\section{Data}
\label{sec:data}
In our analysis, we use images with under 15000 sources from the SKYSURF archival project, detailed in depth in \cite{Windhorst}. This limit removes images with many objects or large galaxies that tend to have double or triple our limit. Using low source count images minimizes the probability that the simulated objects discussed in \S\ref{sec:inserted} are inserted on top of sources already present in the images. This data, from SKYSURF's drizzled WFC3 and ACS/WFC database consisting of 140000 images, is taken from all $HST$ Archive WFC3 and ACS images between 2002 and 2020, with WFC3 images from either the Ultraviolet and Visible Light (UVIS) or Infrared (IR) cameras. All images are drizzled to a resolution of $0\farcs060$ pixel for image consistency, and are from all over the sky. Therefore, the subset of images used in our analysis are very representative of low source count WFC3 Hubble images, and can be used to model the Exposure Time Calculator (ETC) values. We use a subset of between 15 and 30 images per filter, randomly selected from a list of images within an exposure time generally in the range of 100 to 10000 seconds, which ensures a representative model of the SKYSURF data. We will see later that the low number of images per filter is enough to capture main components that affect the completeness.

\section{Analysis}
\label{sec:analysis}

\subsection{Inserted Objects}
\label{sec:inserted}
\begin{figure}[ht!]
\includegraphics[width=1.05\linewidth]{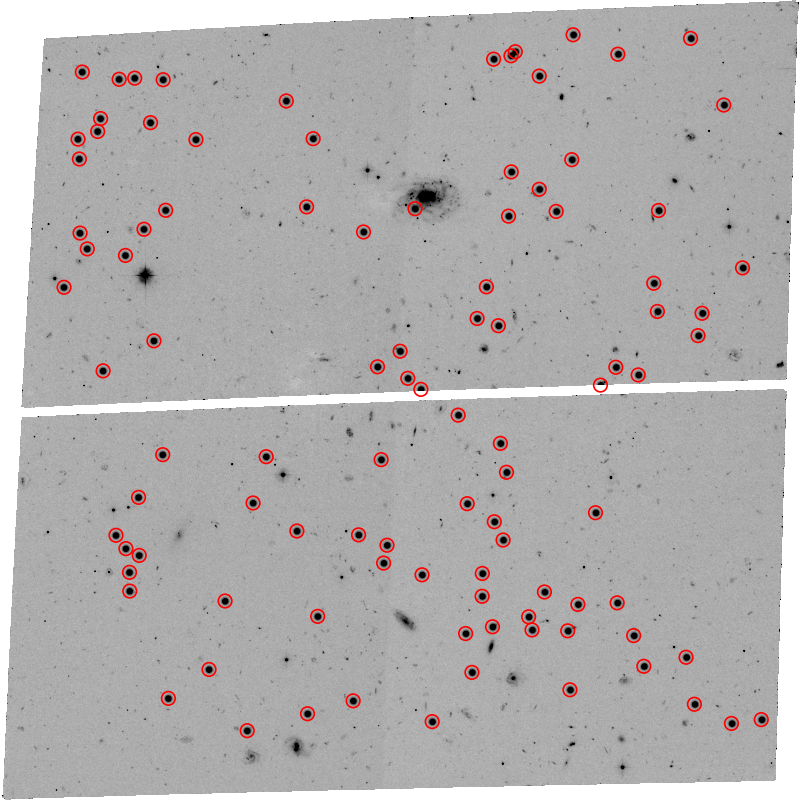}
\caption{An example ACS F475W image showing the 100 inserted objects at 20 AB magnitude and radius of $0\farcs306$.
\label{fig:1}}
\end{figure}

In order to simulate data analogous to the ETC data, we add objects of the same radii and magnitude into a drizzled image, using the GalSim Python package from \cite{Rowe}. As previously mentioned, these shallower drizzled images have $<$15000 sources each to reduce any significant overlap between sources. Our code randomly selects a location in the image away from the edges. To best simulate observed galaxies in the dataset, the objects have a S$\mathrm{\acute{e}}$rsic index of 1 and an inclination of 0\textdegree. Afterwards, we convolve a Gaussian point spread function with a specific flux, dependent on the object's magnitude, and the size of the object, determined by the object's radius. The full-width half-maximum values of the objects are dependent on the filter and determined from the Hubble Ultra Deep Field surveys of \cite{Beckwith} and \cite{Koekemoer}. This process is done 200 pixels from the image edges to avoid drizzling artifacts, and repeated 100 times for each image, to decrease the effect of potential overlap with image sources, as depicted in Figure \ref{fig:1}. We repeat this again for many various sizes, magnitudes, and images, in order to obtain a clearer picture of the completeness of each image.

\subsection{Source Extractor}
\label{sec:sourceExtractorAnalysis}
In order to check whether the sources were detected, we use the Source Extractor program from \cite{Bertin}, which creates a catalog of objects in each image. An object is matched if it is within three pixels of one of our inserted object's locations. Because the used $HST$ images have few sources (under 15000 sources), the impact from overlapping objects, i.e. the confusion limit, is negligible most of the time. The Source Extractor parameters include: a minimum detection area of 5.0, a detection threshold of 1, an analysis threshold of 1, and a 3-sigma Gaussian 5x5 convolution filter. The zero point is determined from the PHOTFLAM and PHOTPLAM components of each image. The number of deblending sub-thresholds is 32 and the minimum contrast parameter for deblending is 0.06.

\section{Results}
\label{sec:results}

\subsection{Fitting}
\label{sec:fitting}
\begin{figure}[ht!]
\includegraphics[width=1.1\linewidth]{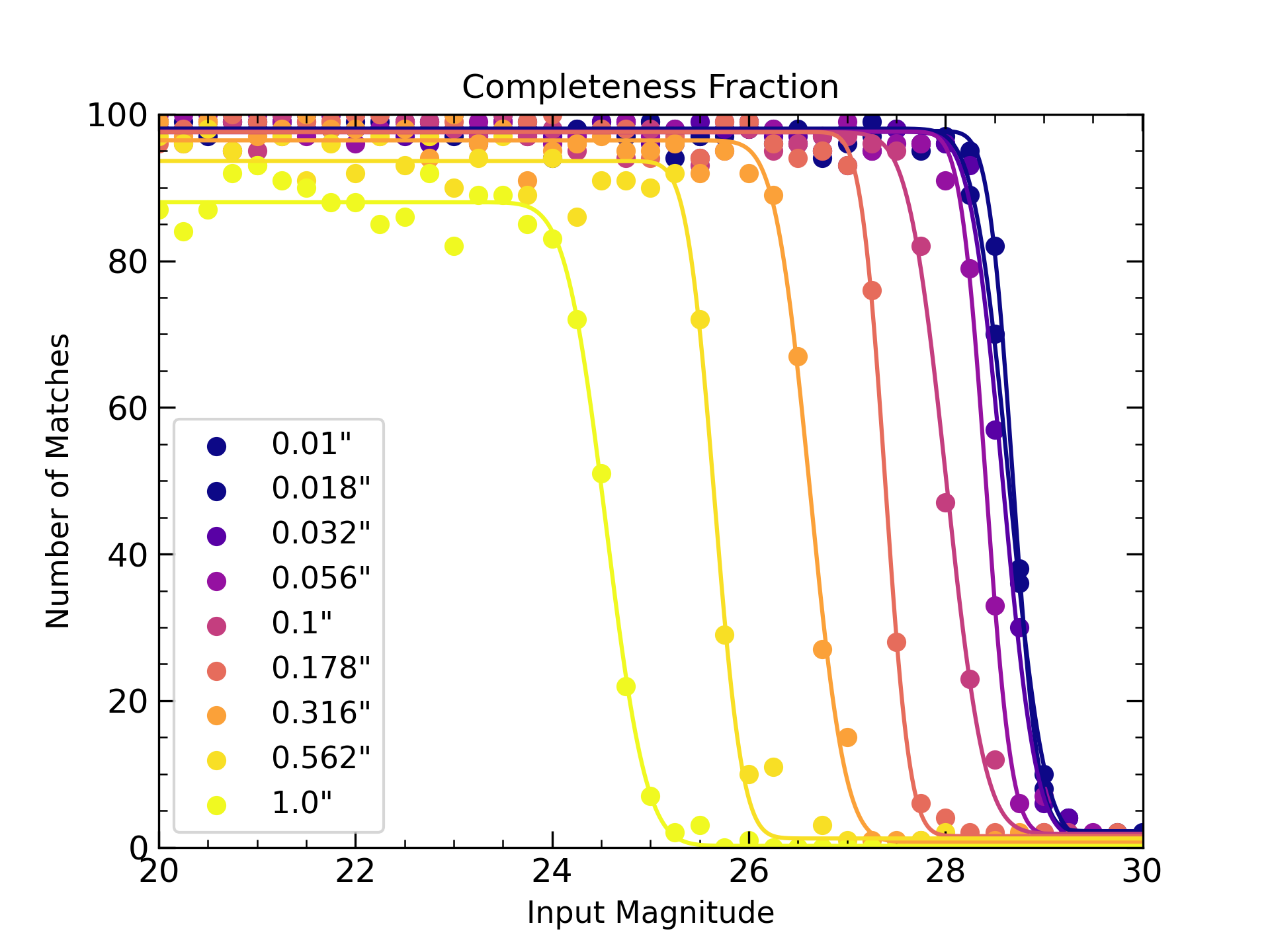}
\caption{The above shows the observed relationship between the completeness limit and magnitude of the objects. Increasing the size of the objects shifted fits to the lower magnitudes at an increasing rate, shown by the color map gradient. This data is also used in Figure \ref{fig:3} to get percent completeness lines. At brighter magnitudes, the confusion limit decreases for larger object sizes due to object overlap, causing fewer matches; however, the general fit remains unaffected. The size and magnitude ranges were chosen as such to represent the majority of observed galaxies, so completeness plots like Figure \ref{fig:3} are representative of many different objects. 
\label{fig:2}}
\end{figure}

\begin{figure}[ht!]
\includegraphics[width=1.15\linewidth]{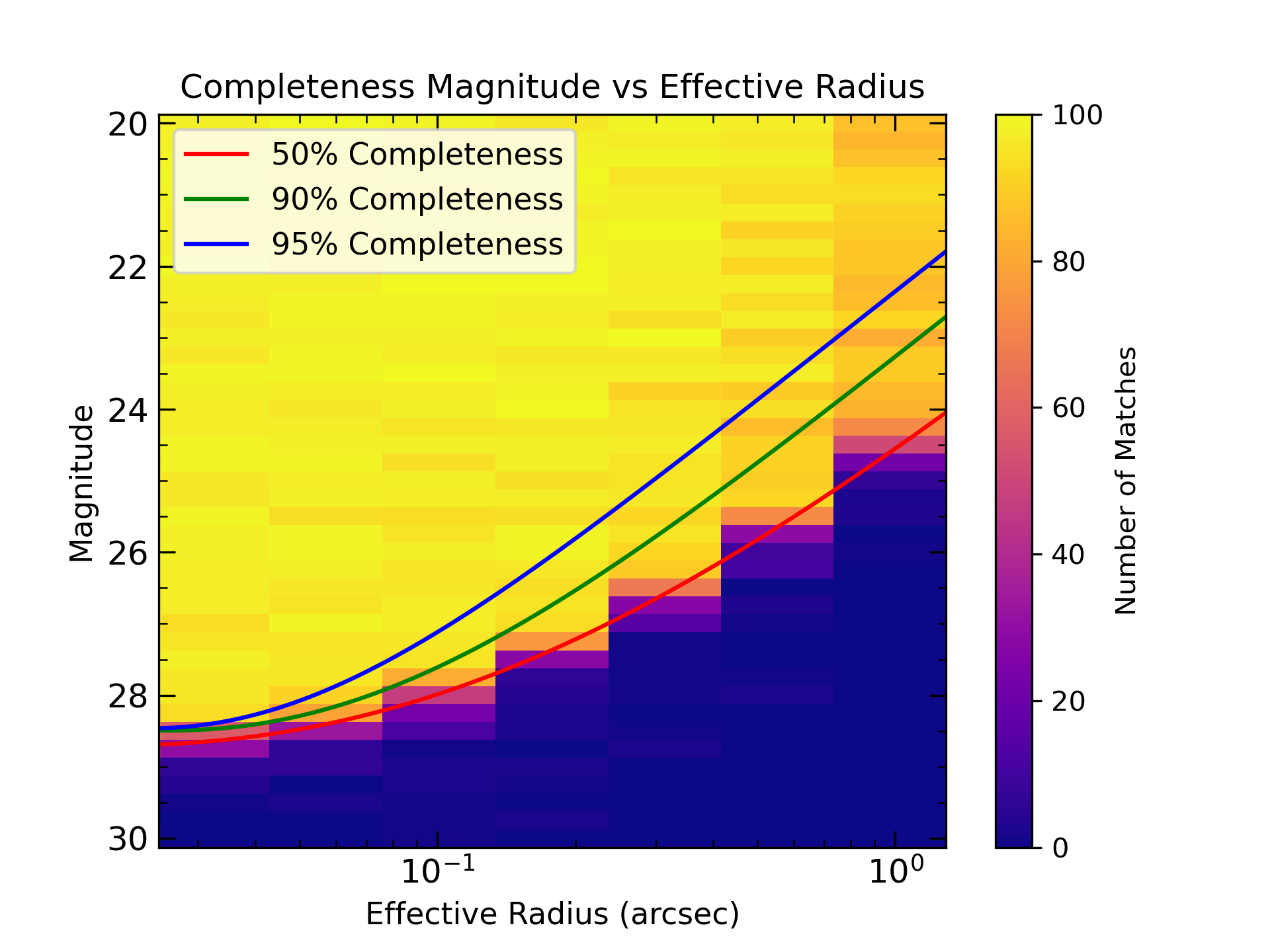}
\caption{Plot showing the completeness limits for a specific object size at each magnitude. The red line follows the line fitted to Equation \ref{equ:1} to the 50\% boxes, illustrating the 50\% completeness for an image. Similarly, the green line corresponds to the 90\% completeness, and the blue line corresponds to the 95\% completeness. Each box represents the number of matches found in an image with inserted objects of the box's magnitude and radius.
\label{fig:3}}
\end{figure}

Using the number of object matches from Source Extractor for each size and magnitude, we construct a set of completeness curves, fit with the complementary error function for each size in every image, which is shown in Figure \ref{fig:2}. 
We create a box plot with a color gradient representing the various number of matches and fit an exponential function to certain completeness levels shown in Figure \ref{fig:3}. The exponential function is the following:

\begin{equation}
y = v + w \frac{\log_{10}(x) - x_0 - 1}{1 + e^{-z(\log_{10}(x) - x_0)}}.
\label{equ:1}
\end{equation}

The four parameters, $v$, $w$, $x_0$, and $z$, are taken from all analyzed images for a given filter and were averaged. Specifically, $v$ is the point source completeness limit, $w$ is the surface brightness slope, $x_0$ is the point source completeness limit magnitude, and $z$ controls the bend from the point source completeness limit. The $x$ variable represents the effective radii of the inserted objects. From this Figure, we can obtain the fit lines for the 50, 90, and 95 percent completeness values, as illustrated in Figure \ref{fig:3}. These are used in \cite{Carter} and follow their independently modeled source counts for each filter in the whole SKYSURF archive.

In Figure \ref{fig:2}, as expected, the completeness percentage decreases when the magnitude gets fainter. The 50\% line in Figure \ref{fig:3} represents the completeness line, where half of the objects at the inserted parameters are detected. For the larger sizes in these two figures, we see the effect of the confusion limit for brighter inserted objects, where the number of matches decreases slightly due to occasional overlap. For a discussion of this effect, see \cite{Kramer}. As the plots show, the effect is not very prominent, only decreasing the maximum number of matches by up to 10\%, so the fits remain largely unaffected. The 0.01\arcsec to 1.0\arcsec object size and the 20 to 30 magnitude ranges capture the bounds of most observed galaxies, so the completeness data is representative of most objects in each filter. Figure \ref{fig:3} takes the points where the number of matches equals fifty, ninety, and ninety-five. 

\subsection{Completeness Model}
\label{sec:Completeness}

\begin{figure}
\centering
\includegraphics[width=1.1\linewidth]{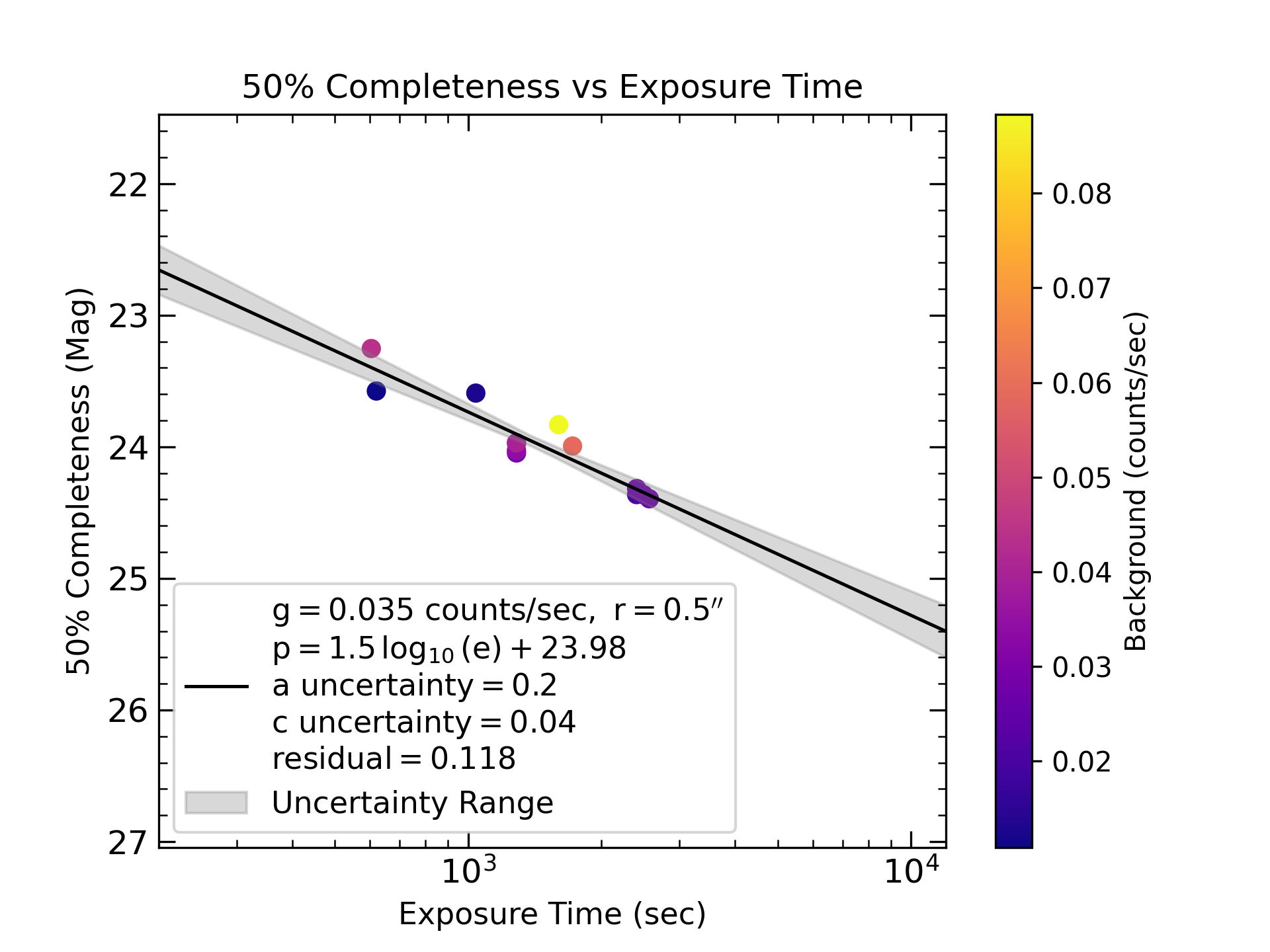}
\caption{Fit of the 50\% completeness magnitude per exposure time for each image in WCF3 UVIS F775W. The color map gradient shows the background values of each image. These variables are combined into a fit shown in blue. The complete figure set for all filters (28 images) is available in the online journal.}
\label{fig:4}

\end{figure}

From the Source Extractor results in Figure \ref{fig:3} using Equation \ref{equ:1}, we take the 50\% completeness limit of each of the images in order to model the completeness for each filter and object sizes 0.1\arcsec, 0.5\arcsec, and 1.0\arcsec. The trend we see here is also present in all other filters. As shown in Figure \ref{fig:4}, we plot a fit modeling the relationship between exposure time and completeness shown in Equation \ref{equ:2} below:

\begin{equation}
p = a (\log_{10}(e) - \langle e \rangle) + c.
\label{equ:2}
\end{equation}

In this equation, $p$ represents the 50\% completeness magnitude, $e$ is the exposure time of the image, $a$ is the exposure time fit parameter, and $c$ is the fit intercept. $\langle e\rangle$ represents the average exposure time and is a filter dependent constant for centering.

This modeling is done independently of the image background values, which ends up having a much smaller impact on the completeness. By looking at the completeness residuals, using the model determined for the specific filter in Figure \ref{fig:4}, the impact of the image background on the completeness is isolated, through Equation \ref{equ:3} below: 

\begin{equation}
R = b (\log_{10}(g) - \langle g \rangle) + f.
\label{equ:3}
\end{equation}

In Equation \ref{equ:3}, the residuals are represented by $R$, the background coefficient by $b$, and the background data values as $g$, with some intercept $f$.  $\langle g\rangle$ represents the average background and is a filter dependent constant for centering. 

\begin{figure}[ht!]
\includegraphics[width=1.1\linewidth]{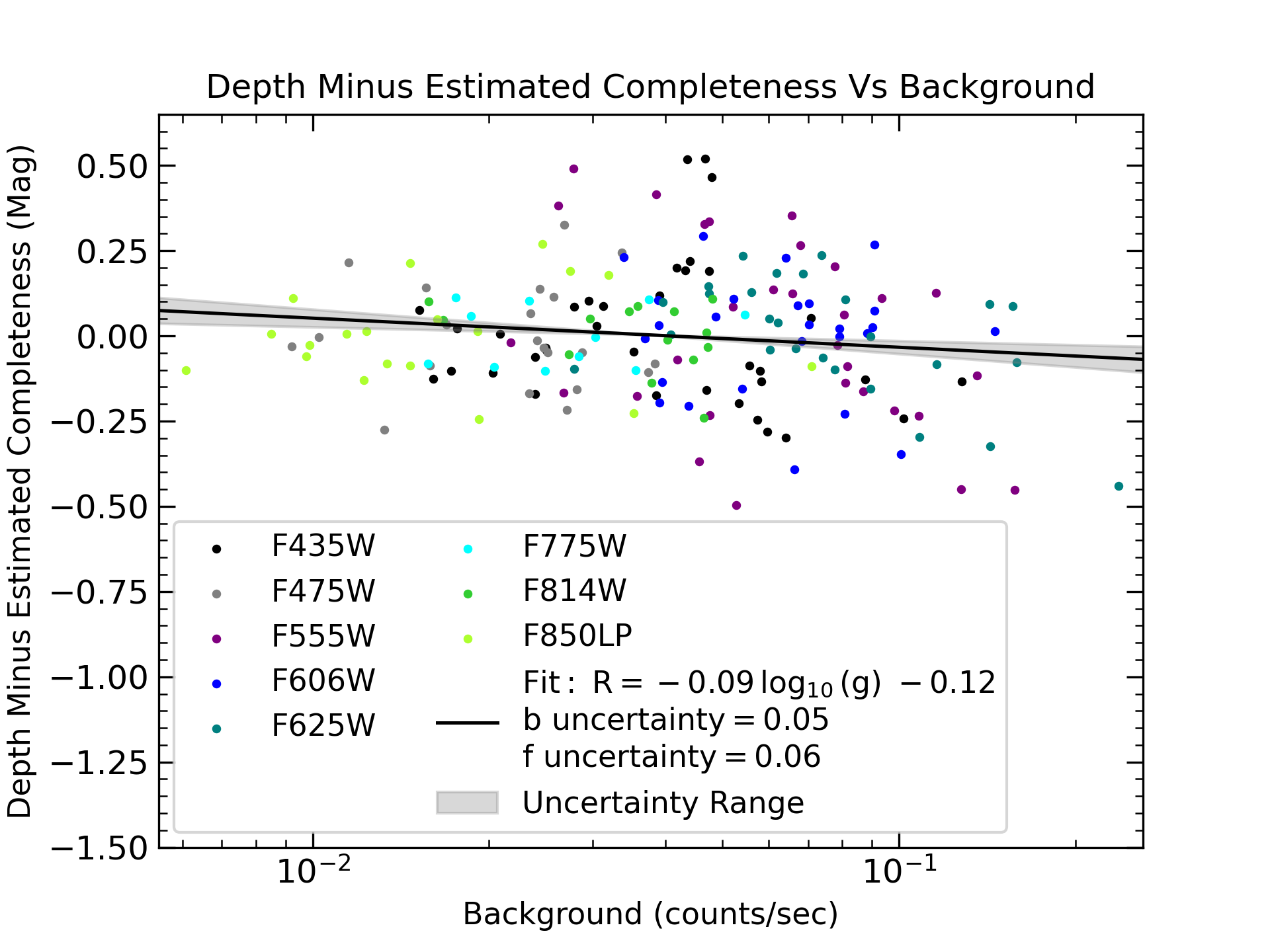}
\caption{Plot illustrating the difference between each individual image's depth against the average depth of the filters for the ACS camera. This helps see the relationship between background level and completeness for different filters.
\label{fig:5}}
\end{figure}

Figure \ref{fig:5} shows the correlation of the background with the actual 50\% completeness values, and is fit for the entire filter set per camera. This slight negative influence of background on our completeness is the same negative dependence discussed before, where a higher background makes it harder to identify fainter objects compared to lower backgrounds.

The exposure time and background models are combined to create our completeness model, shown below in Equation \ref{equ:4}. In Equation \ref{equ:4}, $p$ represents the 50\% completeness magnitude, $e$ is the exposure time of the image in seconds, and $g$ is the background in $\mathrm{e^-/s}$. The constants $a$, $b$, and $c$, are the fit parameters associated with the logarithmic exposure time, the logarithmic background, and the intercept magnitude, respectively.  $\langle e\rangle$ represents the average exposure time and $\langle g\rangle$ represents the average background, which are both filter dependent constants for centering. These constant values are tabulated for all filters in Tables \ref{tab:1}, \ref{tab:2}, and \ref{tab:3} in the Appendix. We define:

\begin{equation}
p = a (\log_{10}(e) - \langle e \rangle) + b (\log_{10}(g) - \langle g \rangle) + c.
\label{equ:4}
\end{equation}


\subsection{ETC}
\label{sec:ETCResults}
Using a logarithmic dependence of magnitude on exposure time and background, we fit the 50\% completeness points for each image at various inserted object sizes to determine the significance of the dependence on each, shown in Figure \ref{fig:3}. These fits can be compared to the ETC data.

In order to obtain the comparable lines from the ETC, we manually record the exposure times for each surface brightness magnitude for each filter and the three object effective radii (0.1\arcsec, 0.5\arcsec, and 1.0\arcsec). To account for varying background levels in the images, we must record the values for all ETC background settings (None, Low, Average, High) and for different object sizes. We also input a signal to noise ratio of 5, extended source with diameter of 8\arcsec, flat continuum in $\mathrm{F_\nu}$, and use AB Magnitude. Lastly, to accurately compare our data to an ETC SNR of 5, we use Equation \ref{equ:5}, where $s$ is the surface brightness, $m$ is the ETC magnitude, and $r$ is the half-light radius of the ETC object. We solve for $s$ using:

\begin{equation}
s = m + 2.5 \log_{10}(\pi r^2).
\label{equ:5}
\end{equation}

\begin{figure}[ht!]
\includegraphics[width=1.1\linewidth]{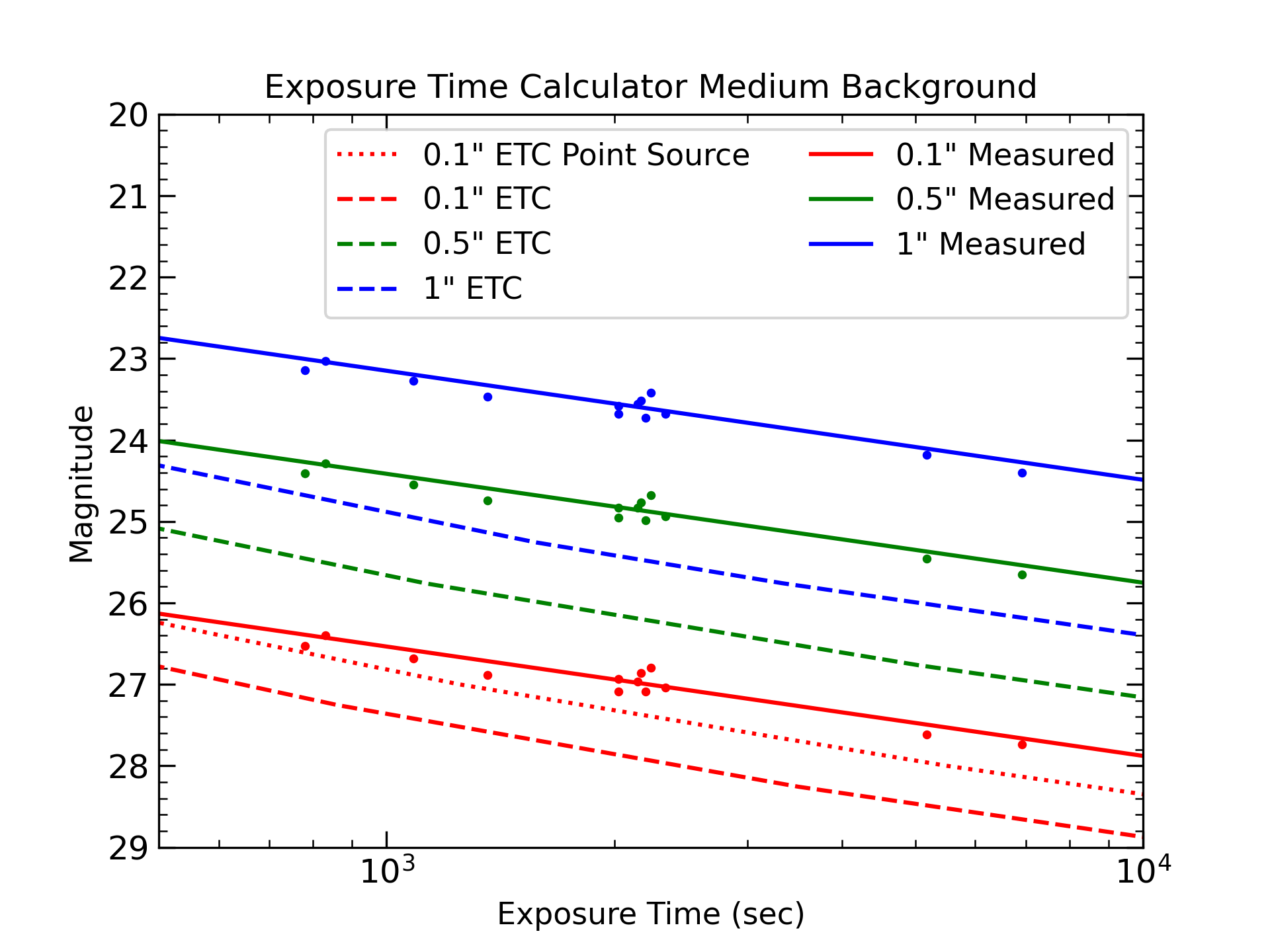}
\caption{Comparison of ACS F814W SKYSURF data to the Exposure Time Calculator for 0.1\arcsec, 0.5\arcsec, and 1.0\arcsec in red, green, and blue, respectively. The ETC values are indicated as a solid line (or a dotted line for point sources). The filled lines represent our SKYSURF measured data. The individual dots are the measured SKYSURF data points from Figure \ref{fig:4} added for reference.
\label{fig:6}}
\end{figure}

Finally, we can compare the two sets of data in Figure \ref{fig:4}. Comparing the ETC values with our plots from Figure \ref{fig:5}, we see a roughly similar trend where the larger sizes and smaller exposures have lower completeness magnitudes. Similarly, the general slope of the lines remains largely the same. The main difference we find, shown by the comparison plot in Figure \ref{fig:6}, is a vertical y-intercept shift, particularly for the larger sizes. We see that the plotted ETC lines for all object sizes end up at a fainter completeness magnitude, up to roughly 1.5 magnitude for 1\arcsec objects. For lower sizes the ETC lines are closer to our data, being around a magnitude fainter.

\section{Discussion}
\label{sec:discussion}
The coefficients for Equation \ref{equ:4}, shown in Tables \ref{tab:1}, \ref{tab:2}, and \ref{tab:3} for each filter, are generally consistent filter-to-filter. However, some exceptions, such as WFC3IR F110W and F140W are noticeable. These high source images may cause the significant deviation seen in these filters. Additionally, we found that the background had less of an impact than the Exposure Time Calculator (ETC) predicts. The UVIS filters had an especially small impact from the background, and we show it as having zero for the $b$ coefficient.

 Although the ETC holds up relatively well and is very useful for exposure time estimations, it is not completely consistent with our findings, shown in Figure \ref{fig:6}. Generally, the 0.1\arcsec individual image values stay within half a magnitude of the ETC; however, the 0.5\arcsec and 1.0\arcsec object sizes vary to even two magnitudes. However, the slope of the our measured lines and the ETC lines appears largely identical. This large discrepancy in magnitude intercept is likely due to the ETC's flat surface brightness model being not good at estimating for our extended sources. 
 
 A GUI calculator of our model is available through the \href{http://skysurf.asu.edu/Completeness/completeness_calculator_gui.html}{SKYSURF website} and the \href{http://skysurf.asu.edu/publiccode.html}{SKYSURF GitHub}, which outputs the 50\%, 90\%, and 95\% completeness limits, given an object's parameters.

\section{Summary}
\label{sec:summary}
In this paper, we measure the completeness limits for a representative set of images from each filter. The completeness is determined by inserting 100 objects of specific magnitudes and sizes, and finding at which magnitude we find desired percentages of the inserted sources using Source Extractor. The 50\%, 90\%, and 95\% magnitude points can be modeled for different object sizes and magnitudes in the different filters. The 50\% completeness limit, in particular, is then modeled for each filter in terms of object size, and exposure and background of the images. 

Ultimately, these results for each filter are compared to the ETC. Generally, we find that the ETC predicts fainter completeness magnitudes for all objects, especially the larger ones. However, the slope of completeness magnitude as a function of the exposure time remains largely identical, leading us to believe that the ETC's flat surface brightness approximation of the sources is the cause of the discrepancy. 

Additionally, our completeness results are a useful independent model to compare to the source counts from the thousands of fields in \cite{Carter}. Furthermore, we plan on using these methods on the James Webb Space Telescope to model its ETC, and to support other future works using the telescope.

\section{Acknowledgments} 
\label{sec:acknowledgements}
Support for program HST AR-15810 was provided by NASA through a grant from the Space Telescope Science Institute. Project SKYSURF is based on observations made with the NASA/ESA Hubble Space Telescope, obtained from the Data Archive at the Space Telescope Science Institute. Dr. Windhorst also acknowledges support from NASA JWST Interdisciplinary Scientist grants NAG5-12460, NNX14AN10G and 80NSSC18K0200 from GSFC.

All of the data presented in this
paper were obtained from the Mikulski Archive for Space
Telescopes (MAST). This project is based on observations
made with the NASA/ESA Hubble Space Telescope and
obtained from the Hubble Legacy Archive, which is a
collaboration between the Space Telescope Science Institute
(STScI/NASA), the Space Telescope European Coordinating
Facility (ST-ECF/ESA), and the Canadian Astronomy Data
Centre (CADC/NRC/CSA).

We thank the anonymous referee for their time and input, as their suggestions greatly improved this paper.

We also acknowledge the indigenous peoples of Arizona,
including the Akimel O’odham (Pima) and Pee Posh
(Maricopa) Indian Communities, whose care and keeping of
the land has enabled us to be at ASU’s Tempe campus in the
Salt River Valley, where this work was conducted.

\textit{Software}: Astropy: \href{http://www.astropy.org}{http://www.astropy.org} \citep{Astropy1, Astropy2}; SourceExtractor: \href{https://www.astromatic.net/software/sextractor/}{https://www.astromatic.net/software/sextractor/} or \href{https://sextractor.readthedocs.io/en/latest/}{https://sextractor.readthedocs.io/en/latest/} \citep{Bertin}; GalSim: \href{https://galsim-developers.github.io/GalSim/_build/html/index.html}{https://galsim-developers.github.io/GalSim/\_build/html/index.html} \citep{Rowe}.

\textit{Facilities}: Hubble Space Telescope Mikulski Archive: 
\href{https://archive.stsci.edu}{https://archive.stsci.edu}; Hubble Legacy Archive (HLA): 
\href{https://hla.stsci.edu}{https://hla.stsci.edu}; Hubble Legacy Catalog (HLC): \href{https://archive.stsci.edu/hst/hsc/}{https://archive.stsci.edu/hst/hsc/}.

\bibliography{eg_udg_references}

\newpage
\begin{table*}[h]
\caption{Table of WFC3 IR completeness coefficients. $p$ represents the 50\% completeness magnitude, and $e$ is the exposure time in seconds and $g$ is the background in $\mathrm{e^-/s}$ of the image. The constants $a$, $b$, and $c$, are the fit parameters associated with the logarithmic exposure time and background, and the intercept magnitude, to be used in Equation \ref{equ:3} below:}
\label{tab:1}
\[
p = a (\log_{10}(e) - \langle e \rangle) + b (\log_{10}(g) - \langle g \rangle) + c.
\]

\centering
\scriptsize
\setlength{\tabcolsep}{3pt}
\renewcommand{\arraystretch}{1.1}
\begin{tabular}{|l| c | c | c | c | c | c | c | c | c | c | c |}
\hline
Filter & 1\arcsec a coeff & 1\arcsec b coeff & 1\arcsec c coeff & 0.5\arcsec a coeff & 0.5\arcsec b coeff & 0.5\arcsec c coeff & 0.1\arcsec a coeff & 0.1\arcsec b coeff & 0.1\arcsec c coeff & $\langle$ e $\rangle$ & $\langle$ g $\rangle$\\
\hline
F098M & 1.3 ± 0.3 & -0.20 ± 0.08 & 24.00 ± 0.06 & 1.3 ± 0.3 & -0.20 ± 0.08 & 25.27 ± 0.06 & 1.3 ± 0.3 & -0.21 ± 0.09 & 27.34 ± 0.06 & 3.583 & -0.391 \\
F105W & 1.3 ± 0.1 & -0.20 ± 0.08 & 24.14 ± 0.05 & 1.3 ± 0.1 & -0.20 ± 0.08 & 25.42 ± 0.05 & 1.3 ± 0.1 & -0.21 ± 0.09 & 27.54 ± 0.05 & 3.418 & -0.135 \\
F110W & 0.73 ± 0.07 & -0.20 ± 0.08 & 24.05 ± 0.03 & 0.71 ± 0.07 & -0.20 ± 0.08 & 25.32 ± 0.03 & 0.65 ± 0.08 & -0.21 ± 0.09 & 27.42 ± 0.03 & 3.028 & 0.079 \\
F125W & 1.1 ± 0.3 & -0.20 ± 0.08 & 24.48 ± 0.05 & 1.1 ± 0.3 & -0.20 ± 0.08 & 25.75 ± 0.05 & 1.1 ± 0.3 & -0.21 ± 0.09 & 27.83 ± 0.06 & 3.708 & -0.195 \\
F140W & 1.4 ± 0.2 & -0.20 ± 0.08 & 24.20 ± 0.05 & 1.5 ± 0.2 & -0.20 ± 0.08 & 25.45 ± 0.05 & 1.6 ± 0.2 & -0.21 ± 0.09 & 27.50 ± 0.06 & 3.499 & 0.021 \\
F160W & 1.0 ± 0.1 & -0.20 ± 0.08 & 24.13 ± 0.05 & 1.0 ± 0.1 & -0.20 ± 0.08 & 25.40 ± 0.06 & 1.0 ± 0.1 & -0.21 ± 0.09 & 27.47 ± 0.06 & 3.645 & -0.189 \\

\hline
\end{tabular}
\end{table*}

\begin{table*}[h]
\caption{Table of WFC3 UVIS completeness coefficients. $p$ represents the 50\% completeness magnitude, and $e$ is the exposure time in seconds and $g$ is the background in $\mathrm{e^-/s}$ of the image. The constants $a$, $b$, and $c$, are the fit parameters associated with the logarithmic exposure time and background, and the intercept magnitude, to be used in Equation \ref{equ:3} below:}
\[
\label{tab:2}
p = a (\log_{10}(e) - \langle e \rangle) + b (\log_{10}(g) - \langle g \rangle) + c.
\]

\centering
\scriptsize
\setlength{\tabcolsep}{3pt}
\renewcommand{\arraystretch}{1.1}
\begin{tabular}{|l| c | c | c | c | c | c | c | c | c | c | c |}
\hline
Filter & 1\arcsec a coeff & 1\arcsec b coeff & 1\arcsec c coeff & 0.5\arcsec a coeff & 0.5\arcsec b coeff & 0.5\arcsec c coeff & 0.1\arcsec a coeff & 0.1\arcsec b coeff & 0.1\arcsec c coeff & $\langle$ e $\rangle$ & $\langle$ g $\rangle$\\
\hline
F225W & 2.4 ± 0.2 & 0.00 ± 0.03 & 22.49 ± 0.04 & 2.5 ± 0.2 & 0.00 ± 0.03 & 23.76 ± 0.04 & 2.5 ± 0.2 & 0.00 ± 0.03 & 25.90 ± 0.05 & 3.36 & 0 \\
F275W & 1.8 ± 0.3 & 0.00 ± 0.03 & 22.19 ± 0.08 & 1.8 ± 0.3 & 0.00 ± 0.03 & 23.47 ± 0.08 & 1.8 ± 0.3 & 0.00 ± 0.03 & 25.63 ± 0.08 & 3.216 & 0 \\
F300X & 1.60 ± 0.07 & 0.00 ± 0.03 & 22.87 ± 0.04 & 1.59 ± 0.08 & 0.00 ± 0.03 & 24.15 ± 0.04 & 1.57 ± 0.08 & 0.00 ± 0.03 & 26.30 ± 0.04 & 3.083 & 0 \\     
F336W & 1.5 ± 0.1 & 0.00 ± 0.03 & 22.89 ± 0.05 & 1.5 ± 0.1 & 0.00 ± 0.03 & 24.17 ± 0.06 & 1.5 ± 0.1 & 0.00 ± 0.03 & 26.33 ± 0.06 & 3.35 & 0 \\
F390W & 0.8 ± 0.2 & 0.00 ± 0.03 & 23.55 ± 0.02 & 0.8 ± 0.2 & 0.00 ± 0.03 & 24.83 ± 0.02 & 0.8 ± 0.2 & 0.00 ± 0.03 & 26.97 ± 0.02 & 3.332 & -1.763 \\      
F438W & 1.5 ± 0.1 & 0.00 ± 0.03 & 23.05 ± 0.06 & 1.5 ± 0.1 & 0.00 ± 0.03 & 24.33 ± 0.06 & 1.6 ± 0.1 & 0.00 ± 0.03 & 26.48 ± 0.06 & 3.375 & -1.843 \\      
F475W & 1.6 ± 0.1 & 0.00 ± 0.03 & 23.88 ± 0.02 & 1.6 ± 0.1 & 0.00 ± 0.03 & 25.17 ± 0.02 & 1.7 ± 0.1 & 0.00 ± 0.03 & 27.34 ± 0.03 & 3.374 & -1.395 \\      
F475X & 1.23 ± 0.08 & 0.00 ± 0.03 & 23.77 ± 0.02 & 1.22 ± 0.08 & 0.00 ± 0.03 & 25.04 ± 0.02 & 1.20 ± 0.08 & 0.00 ± 0.03 & 27.20 ± 0.02 & 3.105 & -1.295 \\  
F555W & 1.5 ± 0.2 & 0.00 ± 0.03 & 23.09 ± 0.05 & 1.5 ± 0.2 & 0.00 ± 0.03 & 24.37 ± 0.05 & 1.5 ± 0.1 & 0.00 ± 0.03 & 26.55 ± 0.04 & 3.07 & -1.419 \\       
F606W & 1.2 ± 0.2 & 0.00 ± 0.03 & 24.05 ± 0.03 & 1.2 ± 0.1 & 0.00 ± 0.03 & 25.33 ± 0.03 & 1.2 ± 0.1 & 0.00 ± 0.03 & 27.51 ± 0.03 & 3.431 & -1.315 \\      
F625W & 1.2 ± 0.1 & 0.00 ± 0.03 & 23.44 ± 0.04 & 1.17 ± 0.09 & 0.00 ± 0.03 & 24.72 ± 0.04 & 1.19 ± 0.09 & 0.00 ± 0.03 & 26.88 ± 0.04 & 3.447 & -1.194 \\
F775W & 1.6 ± 0.2 & 0.00 ± 0.03 & 22.70 ± 0.04 & 1.5 ± 0.2 & 0.00 ± 0.03 & 23.98 ± 0.04 & 1.5 ± 0.2 & 0.00 ± 0.03 & 26.13 ± 0.04 & 3.156 & -1.522 \\       
F814W & 1.5 ± 0.1 & 0.00 ± 0.03 & 22.93 ± 0.05 & 1.5 ± 0.1 & 0.00 ± 0.03 & 24.22 ± 0.04 & 1.6 ± 0.1 & 0.00 ± 0.03 & 26.40 ± 0.04 & 3.252 & -1.293 \\       
F850LP & 1.2 ± 0.2 & 0.00 ± 0.03 & 21.43 ± 0.04 & 1.2 ± 0.2 & 0.00 ± 0.03 & 22.70 ± 0.04 & 1.2 ± 0.2 & 0.00 ± 0.03 & 24.84 ± 0.04 & 3.05 & -2.058 \\ 
\hline
\end{tabular}
\end{table*}

\begin{table*}[h]
\caption{Table of WFC3 ACS completeness coefficients. $p$ represents the 50\% completeness magnitude, and $e$ is the exposure time in seconds and $g$ is the background in $\mathrm{e^-/s}$ of the image. The constants $a$, $b$, and $c$, are the fit parameters associated with the logarithmic exposure time and background, and the intercept magnitude, to be used in Equation \ref{equ:3} below:}
\label{tab:3}
\[
p = a (\log_{10}(e) - \langle e \rangle) + b (\log_{10}(g) - \langle g \rangle) + c.
\]

\centering
\scriptsize
\setlength{\tabcolsep}{3pt}
\renewcommand{\arraystretch}{1.1}
\begin{tabular}{|l| c | c | c | c | c | c | c | c | c | c | c |}
\hline
Filter & 1\arcsec a coeff & 1\arcsec b coeff & 1\arcsec c coeff & 0.5\arcsec a coeff & 0.5\arcsec b coeff & 0.5\arcsec c coeff & 0.1\arcsec a coeff & 0.1\arcsec b coeff & 0.1\arcsec c coeff & $\langle$ e $\rangle$ & $\langle$ g $\rangle$\\
\hline
F435W & 1.2 ± 0.1 & -0.09 ± 0.05 & 23.73 ± 0.03 & 1.2 ± 0.1 & -0.09 ± 0.05 & 25.0 ± 0.03 & 1.2 ± 0.1 & -0.08 ± 0.04 & 27.14 ± 0.04 & 3.463 & -1.412 \\
F475W & 1.4 ± 0.1 & -0.09 ± 0.05 & 23.84 ± 0.04 & 1.4 ± 0.1 & -0.09 ± 0.05 & 25.12 ± 0.03 & 1.5 ± 0.1 & -0.08 ± 0.04 & 27.26 ± 0.03 & 3.303 & -1.669 \\
F555W & 1.6 ± 0.1 & -0.09 ± 0.05 & 23.38 ± 0.06 & 1.5 ± 0.1 & -0.09 ± 0.05 & 24.69 ± 0.05 & 1.3 ± 0.1 & -0.08 ± 0.04 & 26.91 ± 0.04 & 3.336 & -1.21 \\
F606W & 1.0 ± 0.1 & -0.09 ± 0.05 & 24.22 ± 0.04 & 1.0 ± 0.1 & -0.09 ± 0.05 & 25.50 ± 0.04 & 1.09 ± 0.09 & -0.08 ± 0.04 & 27.67 ± 0.03 & 3.499 & -1.208 \\
F625W & 1.3 ± 0.1 & -0.09 ± 0.05 & 23.69 ± 0.03 & 1.3 ± 0.1 & -0.09 ± 0.05 & 24.96 ± 0.03 & 1.3 ± 0.1 & -0.08 ± 0.04 & 27.09 ± 0.03 & 3.388 & -1.119 \\
F775W & 1.12 ± 0.09 & -0.09 ± 0.05 & 23.55 ± 0.03 & 1.13 ± 0.09 & -0.09 ± 0.05 & 24.82 ± 0.03 & 1.2 ± 0.1 & -0.08 ± 0.04 & 26.96 ± 0.03 & 3.309 & -1.583 \\
F814W & 1.34 ± 0.09 & -0.09 ± 0.05 & 23.66 ± 0.03 & 1.34 ± 0.09 & -0.09 ± 0.05 & 24.92 ± 0.03 & 1.3 ± 0.1 & -0.08 ± 0.04 & 27.04 ± 0.03 & 3.349 & -1.459 \\
F850LP & 1.3 ± 0.1 & -0.09 ± 0.05 & 22.97 ± 0.03 & 1.3 ± 0.1 & -0.09 ± 0.05 & 24.23 ± 0.03 & 1.3 ± 0.1 & -0.08 ± 0.04 & 26.36 ± 0.03 & 3.411 & -1.796 \\
\hline
\end{tabular}
\end{table*}

\end{document}